\documentclass[12pt,preprint]{aastex}

\makeatletter

\newenvironment{inlinefigure}{%
\def\@captype{figure}%
\noindent\begin{minipage}{0.999\linewidth}\begin{center}}
{\end{center}\end{minipage}\smallskip}
\makeatother

\def\asca       {{\em ASCA}\/}
\def\chandra    {{\em Chandra}\/}
\def\einstein   {{\em Einstein}\/}
\def\xmm        {XMM-{\em Newton}\/}
\def\rosat      {{\em ROSAT}\/}

\def\mydegree{$^\circ\mskip-5mu$}
\def\myarcmin{$^\prime\mskip-5mu$ }
\def\myarcsec{$^{\prime\prime}\mskip-5mu$}

\usepackage{mathptmx}

\begin{document}

\title{\emph{Chandra} view of the dynamically young cluster of galaxies A1367  \mbox{ }I. small-scale structures}

\author{M.\ Sun \& S.\ S.\ Murray}
\affil{Harvard-Smithsonian Center for Astrophysics, 60 Garden St.,
Cambridge, MA 02138;\\ msun@cfa.harvard.edu}

\shorttitle{\chandra\ view of A1367 I.}
\shortauthors{Sun \& Murray}

\begin{abstract}
The 40 ks \emph{Chandra} ACIS-S observation of A1367 provides new
insights into small-scale structures and point sources in this
dynamically young cluster. Here we concentrate on small-scale
extended structures. A ridge-like structure around the center
(``the ridge'') is significant in the \chandra\ image. The ridge,
with a projected length of $\sim$ 8 arcmin (or 300 h$_{0.5}^{-1}$
kpc), is elongated from northwest (NW) to southeast (SE), as is the
X-ray surface brightness distribution on much larger scales ($\sim$
2 h$_{0.5}^{-1}$ Mpc). The ridge is cooler than its western and
southern surroundings while the differences from its eastern and
northern surroundings are small. We also searched for small-scale
structures with sizes $\sim$ arcmin. Nine extended features, with
sizes from $\sim$ 0.5$'$ to 1.5$'$, were detected at significance
levels above 4 $\sigma$. Five of the nine features are located in
the ridge and form local crests. The nine extended features can be
divided into two types. Those associated with galaxies (NGC 3860B,
NGC 3860 and UGC 6697) are significantly cooler than their surroundings
(0.3 - 0.9 keV vs. 3 - 4.5 keV). The masses of their host galaxies are
sufficient to bind the extended gas. These extended features are
probably related to thermal halos or galactic superwinds of their
host galaxies. The existence of these relatively cold halos imply
that galaxy coronae can survive in cluster environment (e.g.,
Vikhlinin et al. 2001). Features of the second type are not apparently
associated with galaxies. Their temperatures may not be significantly
different from those of their surroundings. This class of extended
features may be related to the ridge. We consider several possibilities
for the ridge and the second type of extended features. The merging
scenario is preferred.

\end{abstract}


\section{Introduction}

Clusters of galaxies exhibit a range of X-ray morphologies from regular and
relaxed to irregular and dynamically young (Forman \& Jones 1982).
Irregular clusters with no X-ray dominant galaxy are generally considered
dynamically young systems. Studies of such dynamically young systems,
in combination with studies of more evolved clusters, are very important
for understanding cluster evolution and its impact on the evolution of galaxies
(and {\em vice versa}). A1367, with its cool gas temperature (3-4 keV; e.g.,
Donnelly et al. 1998; D98 hereafter), high fraction of spiral galaxies,
irregular X-ray distribution and low central galaxy density, is the
prototype of such a dynamically young system (Bahcall 1977; Jones et al. 1979;
Bechtold et al. 1983; B83 hereafter). It is a nearby (z=0.022) X-ray-bright
cluster. Thus, detailed X-ray studies can be performed and the results will
be very important for understanding dynamically young clusters.
 
Based on the \einstein\ HRI (EHRI hereafter) observations, B83 found
fifteen 1$'$ scale emission features in A1367 with luminosities of several
times 10$^{41}$ ergs s$^{-1}$ (0.5 - 4.5 keV). Most features seemed to be
associated with galaxies and were attributed to hot
galactic coronae (B83). However, Canizares, Fabbiano \& Trinchieri (1987)
noted that most early-type galaxies identified in A1367 by B83 have X-ray
luminosities that exceed those of other galaxies with similar optical
luminosities by nearly two orders of magnitude. Thus, they suggested that 
those regions found by B83 might be due to clumpiness in the ICM and should
not be identified with galaxy halos. Grebenev et al. (1995; G95 hereafter)
applied a wavelet analysis to both \rosat\ PSPC and EHRI data. They detected
8 more extended features mainly from the PSPC observation (in a larger field of
view - FOV), though they only detected 8 of 15 extended features claimed by
B83. Lazzati et al. (1998; L98 hereafter) performed their own wavelet
analysis on the same PSPC data. Their results do not show many extended
features ($\sim$ 7) and they only detected 4 of 11 extended features found by
G95. All three analyses found extended features that are not apparently
associated with galaxies though their nature is still vague.

Here we present the analysis of the recent \chandra\ observation of A1367. With
\chandra's superior spatial resolution ($\sim$ 0.5$''$ at the aimpoint) and
better sensitivity than previous missions, we can better understand these
extended features. In this paper, we concentrate on the small-scale structures.
The analysis of point sources (including those corresponding to the member
galaxies) will be presented in paper II (Sun \& Murray 2002). The \chandra\ data
reduction is described in $\S$2; the detection of the extended features is in
$\S$3; $\S$4 describes the \chandra\ temperature map; $\S$5 discusses the
central ridge and a surface brightness discontinuity at its southern end; $\S$6
and $\S$7 discuss extended features; and $\S$8 is the discussion while $\S$9 is
the summary. Throughout this paper we assume H$_{0}$ = 50 km s$^{-1}$
Mpc$^{-1}$ and q$_{0}$ = 0.5. These cosmological parameters correspond to a
linear scale of 0.62 kpc/arcsec at the cluster redshift. All luminosities,
densities and masses scale as h$_{0.5}^{-2}$, h$_{0.5}^{3/2}$ and
h$_{0.5}^{-3/2}$ respectively.

\section{\chandra\ observation \& data reduction}

A1367 was observed on February 26, 2000 by \chandra\ with the Advanced CCD
Imaging Spectrometer (ACIS). The observation was taken by the ACIS-S instrument
using chips 2-3 and 5-8. The roll angle of the observation was chosen
in such a way that the ACIS-S chips lay along the SE-NW elongation of the
cluster. The focal plane temperature was -120\mydegree\ C. The total exposure
was 40.2 ks. The data were telemetered in Faint mode and \asca\ grades 1, 5 and
7 were excluded. Known bad columns, hot pixels, and chip node boundaries were
also excluded. This observation contains three particle background flares that
were excluded based on the X-ray light curves from the outer parts of the field.
The back-illuminated (BI) chips and front-illuminated (FI) chips were filtered
separately since they have different background levels (see Markevitch et
al. 2000 for details). The streak events in the S4 chip were removed by DESTREAK in
CIAO. The effective exposures were 37.8 ks for the S1 and S3 chips, 37.4 ks for
the S2 and S4 chips, and 39.0 ks for the I2 and I3 chips.
Absolute positions of X-ray sources were carefully checked using optical
identifications. A small boresight error of $\sim$ 2$''$ in the original data
was corrected. The derived positions are accurate to $\sim$ 0.5$''$ near the
aimpoint, with larger errors further out.

We generally used local background when we analyzed small-scale structures.
However, the blank field background data relevant to this observation
(Markevitch et al. 2000) were also used in some cases (e.g., temperature map).
The renormalization factors at the time of observation (S1, 3: 1.08; S2, 4:
1.06; I2, 3: 1.11) were applied to the produced background
(http://asc.harvard.edu/\~{}maxim/axaf/acisbg/ ). Since vignetting is dependent on
energy, we divided the 0.5 - 5 keV energy band into six narrow bands and applied
exposure corrections to the narrow-band images. Weighting spectra were also
applied in producing exposure maps (the weighting spectrum differs from chip to
chip based on the spectral fit to the integrated spectrum of each chip).
The ARFs were calculated by weighting the mirror effective area in the region
with the observed cluster brightness distribution in the 0.5 - 2.0 keV energy
band. The spatial nonuniformity of the CCD quantum efficiency (QEU map) was
included in the ARF calculation. An additional position-independent correction
factor of 0.93 for the FI chip quantum efficiency below 1.8 keV was also used to
account for the difference between ACIS-S3 and FI chips (Vikhlinin 2000).
The RMF were calculated by weighting the standard set of matrices within the
region by the observed cluster brightness distribution.

The calibration data used correspond to CALDB 2.9 from \chandra\ X-ray Center
(CXC). The errors in this paper are 90\% confidence intervals unless specified.
The solar abundance table (photospheric measure) by Anders \& Grevesse (1989)
was used. A galactic absorption of 2.2$\times$10$^{20}$ cm$^{-2}$ and the
cluster redshift of 0.022 were used in the spectral fits and computations
unless they are specified.

\section{Detection of extended features}

Previous papers on the analyses of EHRI and \rosat\ PSPC data claimed 26
extended sources in this cluster: B83 - 15, G95 - 16 and L98 - 7 (4
of L98 sources are only candidates). The FWHMs of these sources range from
0.5$'$ to 3$'$ and their luminosities are generally around 10$^{41}$ ergs
s$^{-1}$ in the 0.5 - 5.0 keV band. However, of these 26 sources
only two are detected in all three papers, although G95 analyzed the same
EHRI data as B83 and L98 analyzed the same PSPC data as G95 (see Table 2 for
details). The superior spatial resolution and high sensitivity
of \chandra, in principle, allow us to examine these extended features and
search for new ones. The sensitivity of \chandra\ for detecting extended sources
depends on the assumed source size, the assumed spectrum and the
position in the CCD chips. For a thermal thin plasma spectrum (MEKAL) with 1 keV
temperature and a solar abundance, and for source sizes from 0.5 to 1.5
arcmin in radius, the sensitivity for this observation is 0.7 - 3.5 $\times$
10$^{40}$ ergs s$^{-1}$ (0.5 - 5.0 keV) across the field. Therefore we should
be able to detect all previously reported extended sources if they are in
the field.

\subsection{Methods}

We generated images in the energy bands 0.5 - 2 and 0.5 - 5 keV.
The former was made to match the energy band of \rosat\ while the latter
includes hard X-ray data. All properties of the detected sources
are derived from the 0.5 - 5 keV image, while the 0.5 - 2 keV
image was used to compare with the previous \rosat\ results. Background was
subtracted based on the blank field observations and exposure corrections were
applied. Before searching for small-scale structures, all point sources were
identified (WAVDETECT) and confirmed by comparison with the local point spread
functions (PSFs) generated by MKPSF in CIAO. Point sources were then replaced
by their surrounding averages. 

To detect small-scale structures two methods were applied. The first is
wavelet decomposition (Vikhlinin, Forman \& Jones 1997). Previous application
of this technique on \rosat\ data of the Coma cluster revealed an extended
structure difficult to be found by normal ways (Vikhlinin et al. 1997),
which is later confirmed by \xmm\ observations (Arnaud et al. 2001). The
scales we used correspond to 1, 2, 4, ... 32, 64 times the image pixel size
(3.9344$''$ per pixel; scale 1 - 7 henceforth). The detection threshold
was set at 4.5 $\sigma$ (correponding to 1 false detection in a
512$\times$512 image). No sources were detected on scales 1 and 2, which
means that point sources have been excluded efficiently. On scale 7 there
is only global cluster emission. Some structures appear on scales 3 - 6.

The second method is as follows. Gaussian smoothing was performed to
point-source-extracted images with different $\sigma$ values of 1, 2, 3, ...
8 pixels (3.9344$''$ per pixel). In each smoothed image, the possible
small-scale structures were first picked out by eye. Aperture photometry
was performed in the photon image to check the significance of potential
sources (the definition of S/N is given in Table 1 and the detection
threshold is set at 4 $\sigma$). On the scale of $\sim$ 8$'$ ($\sim$ 300
h$_{0.5}^{-1}$ kpc), a central ridge-like structure (Fig. 1, 2 and 3)
is quite significant. This structure lies along
the SE-NW direction with a dimension of about 3$'\times$8$'$ and its SE end
twists toward the east. On smaller scales, some small structures with sizes
around 1$'$ were revealed. We excluded features that may be related to the
possible residual streak pattern in the S4 chip and the gaps of the CCD chips.
The results of these two methods agree. Nine extended features with sizes from
0.5$'$ to 1.5$'$ were detected.

We measured the position, size, and net count of each feature. Positions were
determined from the average of the peaks of each feature in the smoothed image
and the wavelet reconstructed image (they agree within 5$''$). The radial
profiles of these features were obtained (e.g., Fig. 4). 
These features are all detected in the wavelet reconstructed images on scales
of 5 or less. Thus, the images on scales of 6 and larger can be taken
as the local background. The flattening of the radial profile at some radius
(e.g., Fig. 4) can also be used to estimate the background. Background
levels determined by these two methods are consistent. Then we measured the
50\% encircled energy average radii (EEAR) of these feature to characterize
their extension. When we calculate the significance and net count of each
feature, the radius of the region is chosen to be twice 50\% EEAR. The
results, with uncertainties, are listed in Table 1.

\subsection{The extended features detected by Chandra}

In Table 1, we list the nine extended features detected by \chandra\ using the
methods described in $\S$3.1. Six have been previously detected (Table 2), while
the other three are newly detected in this observation (C2, C5 and C6; see Fig.
2 and 3). Source Gp17 found by G95 may include C2. \chandra\ sources C5 and C6
are too faint to be picked
out by previous observations. There is a detector feature across C6 so that it
is the most poorly determined one. The extended nature of these sources is
presented in Fig. 4, 5, and 9. In Fig. 4, the radial profiles of C1, C7 and C8
are shown. C1 is shown since it lies farthest from the aimpoint, while
C7 and C8 are shown because they are associated with galaxies.
Radial profiles of C2 - C6 are all similar to that of C1. The angular extension
of C1 - C9 vs. PSFs is shown in Fig. 5. The X-ray contours of C9, as well
as the local PSF, are shown in Fig. 9.

In summary, this observation detected 9 extended features on scales of
$\sim$ 1 arcmin in A1367. Three are associated with
galaxies: C7 - NGC 3860B; C8 - NGC 3860; C9 - UGC 6697. Based on the
DSS II image, the other six (C1 - C6) are not associated with
galaxies. There are several member galaxies nearby (e.g., a member galaxy lies
1$'$ east of C1 and another galaxy lies between C2 and C3; see Fig. 2b), but all
have an offset of at least 0.5$'$ (or about 20 kpc) from the centers of C1 - C6.
The galaxies with small boxes in Fig. 2b (right panel) include all known member
galaxies in the field and generally have B magnitudes less than 16.5, but are
not as bright as those associated with C7 - C9. Therefore,
we can divide these 9 extended features into two categories:
features without associated galaxies (C1 - C6) and features
with associated galaxies (C7 - C9). They are discussed
separately in $\S$6 and $\S$7.

\subsection{The extended features detected by Einstein and ROSAT}

We also checked extended features claimed in previous papers (B83, G95 and L98).
It was found that 18 of 26 extended features claimed by previous analyses of
EHRI and \rosat\ PSPC data (B83, G95 and L98) are in the \chandra\ FOV.
Images in the same energy bands as those of EHRI and PSPC were generated. All
18 sources were examined. However, as shown in Table 2, only
6 are confirmed to be extended features in this \chandra\ observation. Four
are point sources or their combination, and 8 are not detected. In the
following we discuss the previous EHRI and PSPC detections separately.

For the EHRI observations, only 2 of 12 extended features reported by B83 and
the same 2 of 8 reported by G95 are confirmed by \chandra. Moreover, 7 of 12
reported by B83 and 3 of 8 reported by G95 (7 in total since 3 are the same)
are not even detected in this \chandra\ observation.
The EHRI sources without \chandra\ detections were also not
detected in the PSPC analyses (G95 and L98). 
For the three EHRI extended features that are actually point sources
or their combination, point sources plus a fluctuation of the cluster
emission can explain the previous results.
For the PSPC observation, 5 of 9 extended features detected by G95
and 3 of 4 detected by L98 are confirmed by \chandra, while
the others are point-like sources or their combination, except Gp27 by G95,
which is not detected by \chandra\ but is also not detected by L98. The
confirmed PSPC extended sources generally lie at small off-axis angles (where
PSFs are still good), while those not detected as extended features are
generally at larger off-axis angles where the PSFs degrade.

In summary, most of the EHRI extended features in the field are not confirmed
by \chandra\ while approximately half of the PSPC extended features in the
field are confirmed. The following factors could contribute to the differences
between the previous results and \chandra\ results: relatively poor photon
statistics of EHRI, unclear systematic uncertainty in the EHRI observations
of A1367, relatively poor spatial resolution of PSPC, and degradation of
PSPC PSFs at large off-axis angles.

\section{The \chandra\ temperature map}

A temperature map was made based on this observation though it covers only a
part of A1367. We divided the field into 42 regions based on photon statistics.
The temperature uncertainty in each region is generally less than 15\%. The
regions with point sources and extended features associated with
galaxies (C7 - C9) were excluded. We fitted the 0.75 - 8.5 keV spectra and
fixed the absorption to the galactic value. A MEKAL model was used and the
abundance was a free parameter in the spectral fit. The results are shown in
Fig. 6.

The \chandra\ temperature map confirms the \asca\ result (D98) that the NW
subcluster has a higher temperature than the SE subcluster. There are temperature
variations across the field. Along the NW-SE direction, temperatures generally
decrease from the NW subcluster to the center of the SE subcluster. The ridge
is located around the coolest part. Temperatures again increase from
the ridge to further SE. Although the \chandra\ temperature map only covers a
part of A1367, it shows several relatively hot regions (15 ,29 and 30) that
may indicate interaction (especially 29 and 30).

\section{The central ridge}

The central ridge is the most striking spatial feature revealed by this
observation. This observation also reveals five extended features (C1 - C5) within
the ridge forming five local crests. The temperature map (Fig. 6) shows that the 
the ridge is located around the coolest part of the cluster.
We also fitted the integrated spectrum of the whole ridge and compared its
temperature with those of the surrounding regions at the east, north,
west and south of the ridge. It was found that the
ridge has a lower temperature (3.2$^{+0.2}_{-0.1}$ keV) than its western
and southern surroundings (3.9$\pm$0.3 keV and 4.1$^{+0.4}_{-0.3}$ keV
respectively), but the same temperature as its eastern and northern surroundings
(3.3$\pm$0.2 keV and 3.5$^{+0.2}_{-0.3}$ keV respectively).

On the scale of about 2 h$_{0.5}^{-1}$ Mpc, the PSPC isophotes can be divided
into two parts: a SE part and a NW part (e.g., D98). Fig. 2a shows that the
outer contours of the SE part are rather spherically symmetrical if we do not
include those elongated to the NW. D98 performed a $\beta$-model fit to this
part (excluding parts contaminated by the NW part). The fit is acceptable but
there is a significant excess around the center. We performed an independent
$\beta$-model fit to
the outer parts of the SE subcluster to examine the central excess. First the
outer part contours are used to determine the center, which is
11$^{\rm h}$44$^{\rm m}$49.$^{\rm s}$8, 19\mydegree42\myarcmin28\myarcsec  (the
cross in Fig. 7). The
regions we chose are from 6.5$'$ to 30$'$ centered at the position mentioned above.
To exclude the contamination from the NW subcluster only the azimuths ranging
from 35\mydegree\ to 225\mydegree\ were included (with the angle measured
counterclockwise from north). The derived core radius is 13.3$\pm$2.2 arcmin and
$\beta$ is 0.84$\pm$0.11. We then subtracted this spherically symmetrical component
from the PSPC image of the whole cluster. The residual (Fig. 7) clearly shows
the central ridge and the NW subcluster. The SE subcluster has been well
subtracted (generally within 3-$\sigma$ fluctuation of the background).
The small-scale features are generally point sources (marked in Fig. 7)
and some extended features. This result implies that the
ridge is a central excess in the SE subcluster.

In Fig. 1, 2 and 3, the surface brightness distribution is relatively
sharp at the southern end of the central ridge, or the south of C1 and
C2. We measured the surface brightness profile across that feature
(Fig. 8). The regions chosen are shown in Fig. 3. Since we intentionally
exclude C1 and C2, the region does not include the whole discontinuity region.
However, the discontinuity feature is significant, as shown in Fig. 8.
It may imply a dynamical process, e.g. a shock or a cold front.
However, current data do not allow us to constrain
temperatures well at each immediate side of the discontinuity.

\section{The extended features without associated galaxies}

Of the nine extended features detected by \chandra, C1 - C6 are especially
interesting since they are not apparently associated with galaxies. C1 - C5
appear to be aligned, forming five local crests along the central ridge. 
The temperatures and abundances of these extended features may help us
understand their nature. Limited by the photon statistics, we fit the spectra
of C1 - C5 together since they are all located within the central ridge.
The spectrum of the region surrounding C1 - C5 was also extracted and
normalized by area. We took it as the background. The best-fit temperature
is 2.3$^{+1.3}_{-0.5}$ keV and the abundance is 0.9$^{+3.0}_{-0.6}$.
If the spectrum of the surrounding was normalized by the predicted value
from the $\beta$-model fit (D98), the results change little. The
spectrum of the surrounding region was also fitted. The best-fit temperature
and abundance is 3.4$\pm$0.2 keV and 0.41$^{+0.10}_{-0.09}$ respectively.
Thus, current
data do not imply any significant difference in temperature between the
extended features and their surroundings. The spectrum of C6 does
not suggest a soft component like those in C7 - C9 ($\S$5) but the current 
data do not allow us to constrain the temperature well.

Next we attempt to estimate the physical parameters of those features. For C1
- C5, the spectrum of each feature was obtained, and fitted by a MEKAL
model with the temperature and abundance fixed at the best-fit values
obtained above (2.3 and 0.9 respectively). From the normalization obtained, 
we can derive the mean proton density and the gas mass assuming a
constant-density sphere. For C6, we simply assumed the same spectrum as
that of C1 - C5. The results, as well as the uncertainties, are listed
in Table 1.

Thus, the statistics of the current data are not sufficient to constrain
the physical parameters of C1 - C6. The integrated C1 - C5 spectrum implies
a temperature around 2.5 keV (with about 1 keV error), not significantly
different from their surroundings. Their X-ray luminosities are all around
10$^{40}$ - 10$^{41}$ ergs s$^{-1}$. The average proton densities and the gas
masses are $\sim$ 10$^{-3}$ cm$^{-3}$ and 10$^{9}$ M$_{\odot}$ respectively.

\section{The extended features associated with galaxies}

Three extended features were found to be associated with galaxies:
NGC 3860 (C8), NGC 3860B (C7) and UGC 6697 (C9) as shown in Fig. 2b and 9. Their
properties are listed in Table 3. Seven other member galaxies detected as
point-like sources are discussed in paper II.

\subsection{Sa galaxy - NGC 3860}

This \chandra\ observation reveals two X-ray components in NGC 3860
(C8), a bright central source and a faint diffuse extension (Fig. 4). The
central source is unresolved and its X-ray luminosity is about 10
times larger than that of the diffuse extension. The detailed analysis of this
point-like source will be addressed in paper II.
Only $\sim$ 110 counts were collected from the diffuse component during
the observation. Its spectrum was extracted. The surroundings were taken as
the background. The net spectrum can be fitted well by a MEKAL model with a
temperature 0.7$\pm$0.2 keV and the luminosity is $\sim$ 1.3$\times$10$^{40}$
ergs s$^{-1}$ (0.5 - 5 keV).

This soft component may be the thermal halo of NGC 3860.
Thermal halos in spiral galaxies with temperatures from 0.2 to 0.6
keV have been reported in several nearby spirals based on \rosat\ observations
(M83 by Ehle et al. 1998; NGC 4258 by Vogler \& Pietsch 1999; M81 by Immler
\& Wang 2001). The luminosities reported in those sources are similar to
that of NGC 3860 and the temperatures are consistent. The observed X-ray
luminosity of the soft component in NGC 3860 also agrees well with the
predicted one scaled from its optical blue luminosity based on L$_{\rm X}$
- L$_{\rm B}$ correlation of field galaxies (e.g., Brown \& Bregman 1998).

\subsection{Starburst galaxies - NGC 3860B \& UGC 6697}

NGC 3860B is a starburst galaxy, based on its bright H$\alpha$ emission and
irregular morphology (private communication with S. Sakai). It appears extended
in this observation with a diameter of $\sim$ 0.5$'$ (Fig. 4). The peak of
the X-ray emission is about 0.1$'$ SW of the optical center. Only about 120
net counts from this source are collected during the observation. If the
surroundings were taken as the background, the derived temperature is
0.5$\pm$0.2 keV with the abundance fixed at solar. The derived L$_{\rm x}$
is $\sim$ 3.0$\times$10$^{40}$ergs s$^{-1}$ (0.5 - 5 keV).

UGC 6697 is a luminous peculiar galaxy with an asymmetrical optical appearance.
H$\alpha$ and ultraviolet observations revealed that it was a starburst
galaxy (Kennicutt et al. 1984; Donas et al. 1990). A radio ``trail'' 
on the NW side was detected by Gavazzi \& Jaffe (1987). It has been suggested
that UGC 6697 is undergoing a dynamical interaction with the ICM assuming
the galaxy is moving toward the SE (Gavazzi et al. 1995). Fruscione \& Gavazzi
(1990) also reported X-ray emission from this galaxy and attributed it to
discrete X-ray sources associated with young stars.

In this \chandra\ observation, UGC 6697 is located in the S1 chip and is
$\sim$ 17$'$ from the aimpoint (the 50\% EEAR of PSF there is $\sim$
17$''$). However, as shown in
Fig. 9, it is clearly resolved. The X-rays follow the optical light and
the SE part is brighter than the NW part. About 800 counts from this source were
collected during this observation. Its spectrum was extracted and the background
was taken from the surroundings. A thermal thin plasma (MEKAL) model can fit the
data well (T=0.70$^{+0.14}_{-0.17}$ keV; $\chi^{2}$=15.7/20) but the abundance is
very low ($<$ 0.1, 3$\sigma$). Additional components, like a power law or another
MEKAL, do not improve the fit significantly. The low heavy element
abundance found in UGC 6697 is probably artificial, related to the relatively
poor photon statistics. The derived X-ray luminosity
is about 1.8$\times$10$^{41}$ ergs s$^{-1}$ (0.5 - 5 keV), which makes
UGC 6697 an X-ray luminous starburst galaxy.

Both spectra of NGC 3860B and UGC 6697 show thermal components with temperatures
of 0.3 - 0.9 keV and their luminosities roughly scale with their L$_{\rm B}$.
Such thermal components with similar temperatures were also reported by Zezas,
Georgantopoulos \& Ward (1998) for NGC 3310 and NGC 3690 ($\sim$ 0.8 keV),
and by Dahlem, Weaver \& Heckman (1998) for some nearby starburst galaxies
(0.65 - 0.9 keV). The observed X-ray to optical luminosity ratios of NGC 3860B
and UGC 6697 also agree with those of nearby starbursts (e.g., Dahlem et al. 1998).
It has been suggested that during the starburst phase
galaxies emit X-rays much more intensely than normal galaxies.
The possible explanation of the thermal components involves a galactic
superwind, integrated emission from supernova remnants and integrated
emission of the young stars (e.g., Rephaeli, Gruber \& Persic 1995).

\section{Discussion}

The \chandra\ observation of A1367 shows complicated structures around its
center: a ridge-like structure and some extended features. For features
not associated with galaxies, even in the absence of thermal evaporation
losses, ram pressure stripping and Rayleigh-Taylor instabilities may destroy
them within 10$^{7}$ - 10$^{8}$ yrs (Nittmann, Falle \& Gaskell 1982; Sarazin
1986). We present several possibilities for the nature of these features.

\vspace{0.5cm}
\noindent
a) Background clusters\\

One may argue that some features are background clusters at sufficiently
high redshift that the galaxies are too faint to show up in a shallow
optical exposure. However, based on the log N - log S relation
for clusters (e.g., Rosati et al. 1995; Kitayama, Sasaki \& Suto 1998),
we expect to see approximately 0.2 clusters at flux levels higher than
1.0$\times$10$^{-14}$ ergs s$^{-1}$ cm$^{-2}$ (0.5 - 2 keV) in the
10$'\times$10$'$ field (the flux of the faintest extended feature - C5 -
is about 1.4$\times$10$^{-14}$ ergs s$^{-1}$ cm$^{-2}$).
Thus, the probability of seeing two (or three) clusters when
0.2 are expected is 1.6$\times$10$^{-2}$ (or 1.1$\times$10$^{-3}$).
Moreover, five of six features (C1 - C5) seem to be located in the ridge and
are aligned with each other, rather than distributed randomly as expected
for background clusters. Thus, we conclude that the possibility for one or more
features to be background clusters is very small.

\vspace{0.5cm}
\noindent
b) Features with no galaxies sustained by dark matter halos with little light\\

C1 - C6 can be supported by massive dark matter halos corresponding to very
faint galaxies. The masses needed to sustain gas clouds from thermal diffusion
can be estimated as the following. The assumed dark matter halos are required
to have masses high enough that GMm$_{\rm H}$/R $>$ 3/2 kT, where R and T are
the sizes and the temperature of the feature respectively. Thus,
M $>$ 6.7$\times$10$^{11}$ (T/1 keV) (R/20 kpc). First we check the binding
masses needed for C7 - C9. Since we know there are member galaxies associated
with them, their total masses can be estimated by assuming M/L =
7(M/L)$_{\odot}$. The results are presented in Table 3. It is clear that
the total masses of the three galaxies are all larger than
the masses needed to bind their corresponding extended features.
We estimated the binding masses needed for C1 - C6. The results are
1 - 3 $\times$10$^{12}$ M$_{\odot}$, which is $\sim$ 10 times larger than
the total mass of any member galaxy close to C1 - C6 unless the nearby member
galaxies have a large mass-to-light ratio of $\sim$ 50. However, the alignment
of C1 - C5 is not easy to explain and their similar temperatures to their
surroundings imply that they are more likely to be inhomogeneities in the
cluster medium.

\vspace{0.5cm}
\noindent
c) Recently stripped galaxy halos\\

Takeda, Nulsen \& Fabian (1984) suggested some galaxyless extended
features reported by B83 might be the blobs of the stripped gas.
Based on the results in Nittmann et al. (1982), they estimated that such
blobs can survive for about 10$^{8}$ years before being destroyed by
Rayleigh-Taylor instabilities. Toniazzo \& Schindler (2001)
pointed out that these gas halos would be accelerated radially by the cluster
potential gradient. Based on the expression they gave for the distance
traveled by the gas blobs, we estimate that the blobs will travel about 50 -
150 kpc radially inward before being destroyed. Within 50 - 150 kpc of C1 - C6,
there are not many known member galaxies (Fig. 2b). Moreover, it is known from
the above discussion that it is difficult to explain each of C1 - C5 by gas
stripping of only one galaxy. This scenario also has problem to explain
the alignment of C1 - C5.

\vspace{0.5cm}
\noindent
d) The fragmented core of a merging component\\

As shown in Fig. 2, on the scale of about 2 h$_{0.5}^{-1}$ Mpc, the X-ray
emission is largely flattened along the SE-NW direction and it can be divided
into two parts spatially: one larger part at the SE near the peak of the cluster
X-ray emission, and the other smaller part at the NW where there is a clear
galaxy concentration centered at NGC 3842. Nevertheless, the galaxy
distribution in the SE does not show any condensation near the X-ray center.
D98 performed $\beta$-model fits to these two components and constructed a
temperature map based on \asca\ data. They suggested an ongoing merging along
the SE-NW direction with two subclusters corresponding to the two parts mentioned.
This \chandra\ observation reveals a ridge around the center elongated in
the same direction as that of the flattened X-ray emission on a scale of about
2 h$_{0.5}^{-1}$ Mpc. Five of six extended features with no associated
galaxy (C1 - C5) lie along the ridge and form five crests within it, while C6 is
still close to the ridge. The ridge may have originally been the core of a
subcluster. It is elongated along the merging direction because of the
interaction with another subcluster. Instabilities (e.g., Kelvin Helmholtz
instability) make it fragment into clumps we now
observe (C1 - C5, possibly C6). The discontinuity found at the southern end
of the ridge suggests motion of the ridge toward the SE. If this is
true, the ridge could not be the core of the SE subcluster since its core
is supposed to move to the NW in the interaction with the NW subcluster.
Moreover, the shape of the ridge implies it has been distorted to a very large
extent, by ram pressure stripping or tidal forces. This is possible
for an infalling gas cloud, but not likely for the core of the larger SE
subcluster at the very early stage of merging. It is difficult to relate the
ridge with the NW subcluster since they are separated spatially (Fig. 2 and 7)
and the NW subcluster peaks around a galaxy concentration centered at NGC 3842.
Therefore, the ridge would have to be a new merging component, perhaps a compact
precursor of the NW subcluster penetrating the SE subcluster. When it passes the
SE subcluster, it is elongated and fragmented by ram pressure stripping or tidal
force. Simulations show that at the early stage of cluster evolution,
clusters may accrete smaller subclusters along filaments (summarized by
Bertschinger 1998). Similar examples in X-ray and optical are presented by
West (1998) for Coma and Durret et al. (1998) for A85. More observations in
wider fields will be helpful to clarify the nature of the ridge.

\section{Summary}

1) A ridge-like structure was detected around the center of the cluster.
The ridge, with a projected size of $\sim$ 3$'\times$8$'$, is elongated along
the SE-NW direction similar to the cluster X-ray emission on scales of $\sim$
2 h$_{0.5}^{-1}$ Mpc. It is cooler than its western and southern surroundings.
A discontinuity of surface brightness was found at the southern end of the
ridge. Although the statistics of the current data do not allow us to constrain
the temperature change across this discontinuity, it suggests a motion of the
ridge toward the SE. We prefer the ridge is a separate merging component,
like a precursor of the NW subcluster.

2) Nine extended features (C1 - C9) were found. They can be
divided into two types. Features of the first type (C7 - C9)
are associated with galaxies NGC 3860, NGC 3860B and UGC 6697 respectively.
Each of their spectra can be fitted well by a single MEKAL model with
a temperature of 0.3 - 0.9 keV, while the surrounding ICM is much hotter
with temperatures of 3 - 4.5 keV. They are probably thermal halos
(especially for C8) and galactic superwind (especially for C7 and C9) of their
host galaxies. This implies that thermal halos can survive in cluster
environment.

3) Features of the second type are not obviously associated with
galaxies. C1 - C5 are aligned and form five crests within the ridge.
The temperatures of these features are not significantly different from
those of their surroundings. Their luminosities are around 10$^{41}$ ergs
s$^{-1}$ (0.5 - 5.0 keV) and the gas masses are several times
10$^{9}$ M$_{\rm \odot}$. Unlike features of the first type,
the masses needed to bind them are about 10 times larger than the masses of any
nearby member galaxy. In view of their tight spatial correspondence with the
central ridge, we suggest they are fragmented clumps of the ridge.

4) The \chandra\ temperature map confirms that the NW subcluster is relatively
hotter than the SE subcluster. It shows some temperature variations and
several interesting relatively hot regions.

5) Previously claimed ``extended features'' were checked. For 18 previously
claimed ``extended features'' in the field, only 6 were confirmed in this
observation. The others are either point-like sources, or their combination,
or not detected. We conclude that previous observations, especially the one
by EHRI, overestimate the number of extended features in this cluster
because of the limitations of those instruments.

\acknowledgments

The results presented here are made possible by the successful effort of the
entire \emph{Chandra} team to build, launch, and operate the observatory. We
are very grateful to W. Forman for valuable comments to the manuscript. We
are also grateful to the referee for the valuable comments. We
would also like to thank A. Vikhlinin to provide his software for data
analysis. We thank K. Rines and P. Berlind for providing spectroscopy
information of some galaxies. We acknowledge helpful discussions with P.
Buckley, G. Fabbiano, J. Girash, J. Huchra and M. Markevitch.
This study was supported by NASA contract NAS8-38248.

\clearpage

\begin{figure}
\vspace{-2cm}
  \centerline{\includegraphics[height=1.3\linewidth]{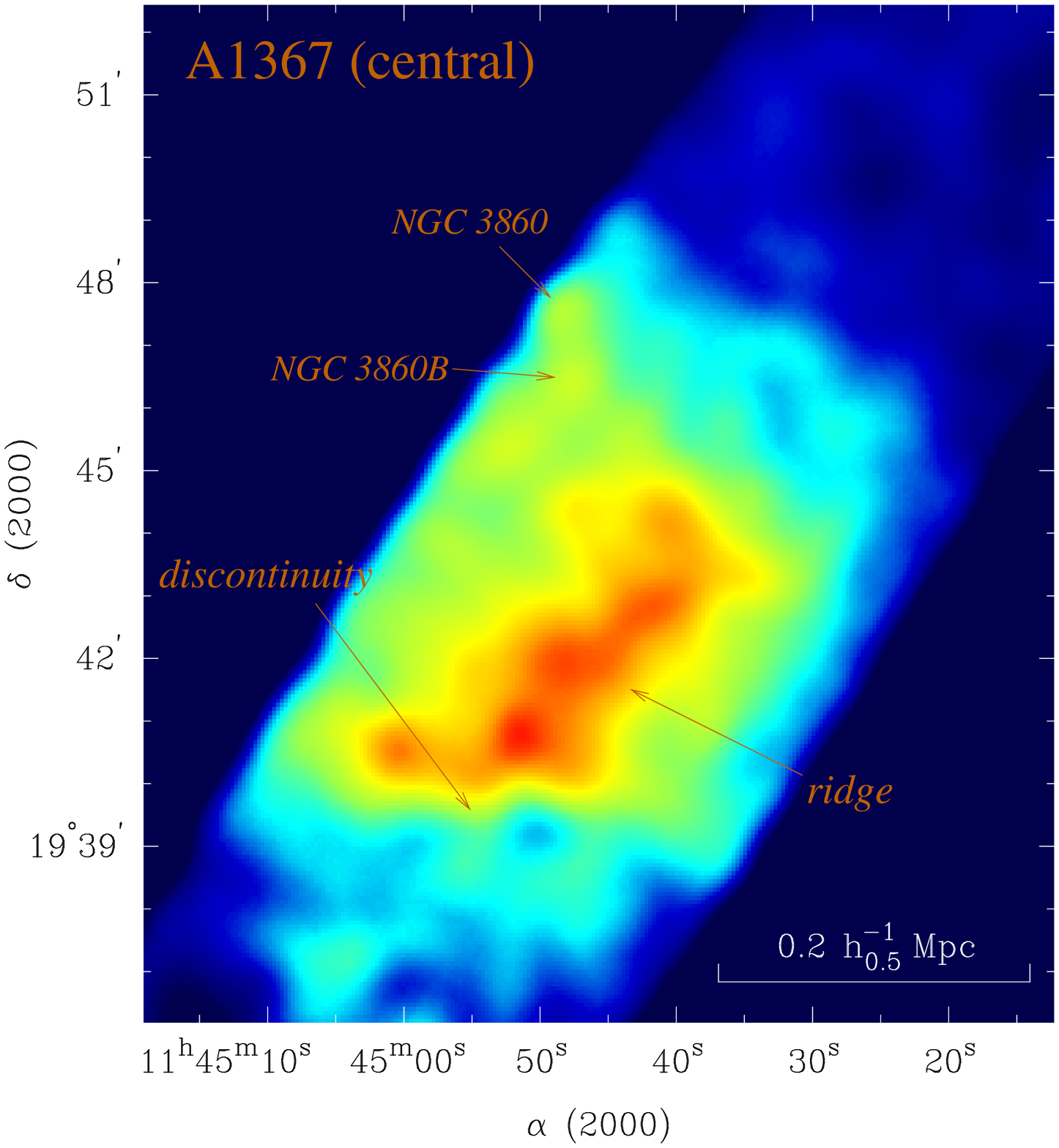}}
\vspace{-2.7cm}
  \caption{The X-ray central part of A1367 viewed by \chandra. The image is
background-subtracted and exposure-corrected in the 0.5
- 5 keV energy band, binned to 3.9344$''$ per pixel and smoothed with a variable
gaussian with 20$'' \sigma$ at the center and 30$'' \sigma$ at the outer
parts. All point sources have been replaced by the surrounding averages.
The central ridge is significant and shows some internal structures.
    \label{fig:img:smo}}
\end{figure}

\clearpage

\begin{figure*}[htb]
\vspace{-6cm}
  \centerline{\includegraphics[height=1.5\linewidth]{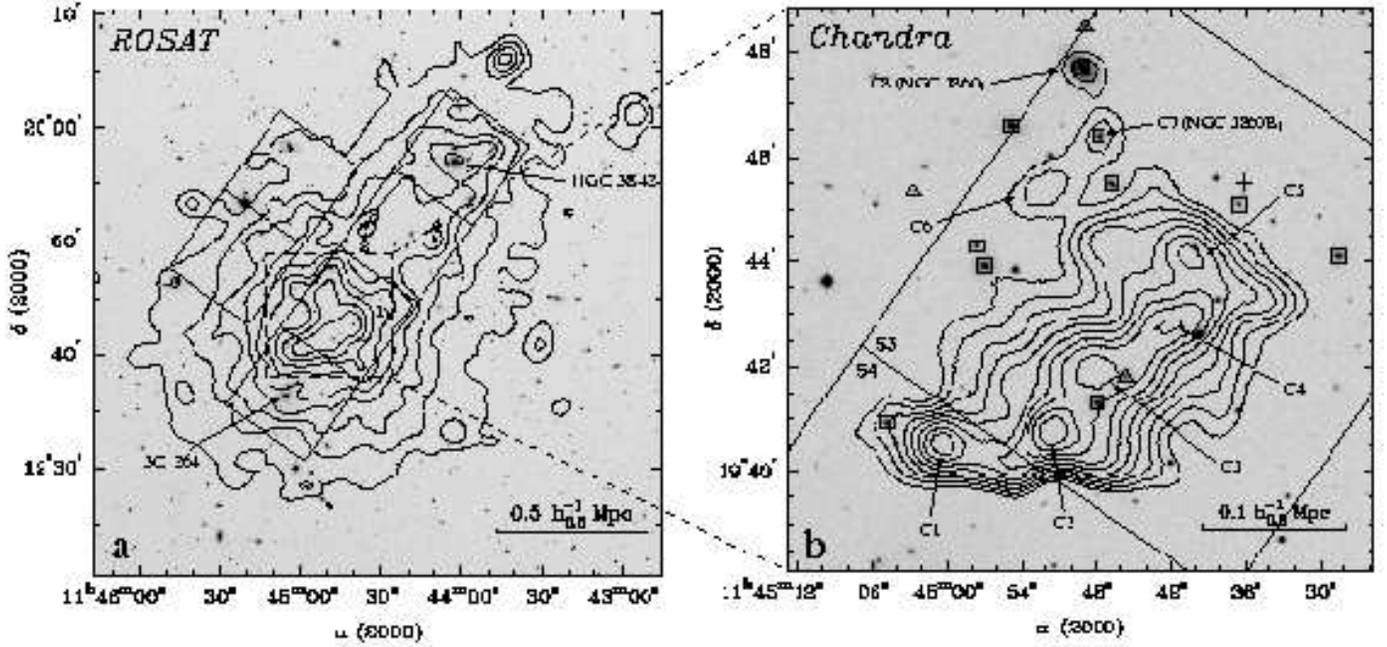}}
\vspace{-7.8cm}
  \caption{a): \rosat\ PSPC contours overlaid on the DSS I image. The two
X-ray brightest point-like sources, 3C 264 and the quasar at the west of NGC
3842 (EXO 1141.3+2013), were excluded. The PSPC image was smoothed by a
gaussian with 45$'' \sigma$. The contour levels are linearly from 25\% to 90\%
of
the maximum. The SE-NW elongation on the scale of $\sim$ 2 h$_{0.5}^{-1}$ Mpc is
significant. The field of the \chandra\ observation is shown by the solid lines.
Some sources detected by \chandra\ are also drawn for the purpose of this
paper, e.g., those claimed to be extended features previously but turns out to
be point sources (1-4; see Table 2).
The dashed-line box delineates the zoom-in central field on the right.
b): The zoom-in central region of A1367 (\chandra\ contours with
DSS II image). The X-ray contours are made from the
smoothed image that the background has been subtracted and the exposure
has been corrected. All X-ray point sources have been excluded.
We used a gaussian with 20$'' \sigma$ to smooth the image.
The contour levels are linearly from 60\% to 95\% of the maximum.
The central ridge is significant, with a projected length of about 300 kpc.
It is also elongated along the same SE-NW direction as that of X-ray emission
on the scale of about 2 h$_{0.5}^{-1}$ Mpc (shown in the left one). The surface 
brightness
distribution at the southern end of the ridge is sharper than those of the
other directions ($\S$6). Eight extended features (C1 - C8) are detected
around the center (C9 is out of the field).
C1 - C5 form five crests of the central ridge. C7 and C8
are associated with member galaxies NGC 3860B and NGC 3860 respectively.
The small boxes represent known member galaxies in the field, while
the small triangles represent known background galaxies. The cross near C5
represents the aimpoint.
    \label{fig:img:smo}}
\end{figure*}

\clearpage

\vspace{4mm}
\begin{inlinefigure}
 \centerline{\includegraphics[height=0.8\linewidth,angle=270]{f3.ps}}
\vspace{1mm}
  \caption{The central part of A1367 by \chandra\ (the gray image - DSS II;
contours - \chandra). The X-ray contours are from the reconstructed
image by wavelet decomposition tool. This figure is shown here for comparison
with Fig. 2b. For convenience, the contours affected by the edges of the chips
are not shown. All features are above 4.5$\sigma$. The contour levels
are linearly
from 40\% to 95\% of the maximum. This reconstructed image is quite similar to
the smoothed one in Fig. 2b. The central ridge is significant. The
substructures revealed are also shown in Fig. 2b (C1 - C8).
The sharp edge at the south is also significant. The regions used in
$\S$5 to measure the surface-brightness profile are also shown here (Fig. 8).
    \label{fig:img:smo}}
\end{inlinefigure}

\clearpage

\begin{figure*}[htb]
  \centerline{\includegraphics[height=1.0\linewidth,angle=270]{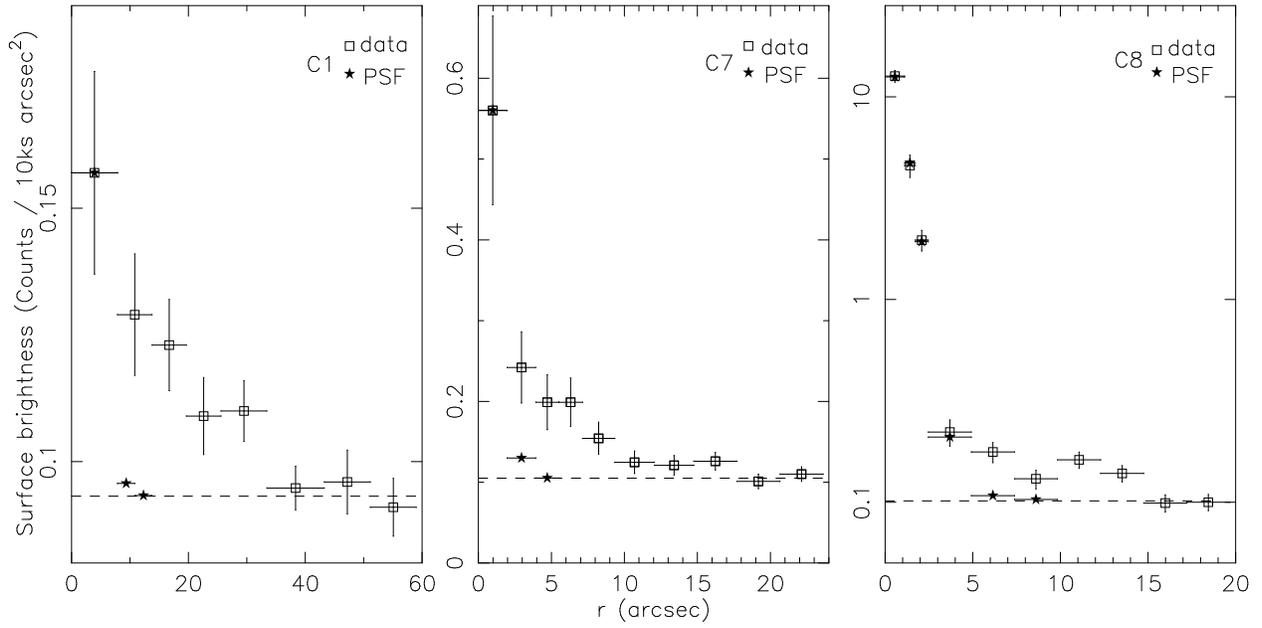}}
  \caption{Radial profiles of extended features C1, C7 (NGC 3860B) and C8
(NGC 3860) with local PSFs. The PSFs were generated by MKPSF in CIAO and
normalized to match the central brightness of C1, C7 and C8. The dashed lines
represent the surrounding background levels.
    \label{fig:img:smo}}
\end{figure*}

\clearpage

\begin{figure}
  \centerline{\includegraphics[height=1.0\linewidth,angle=270]{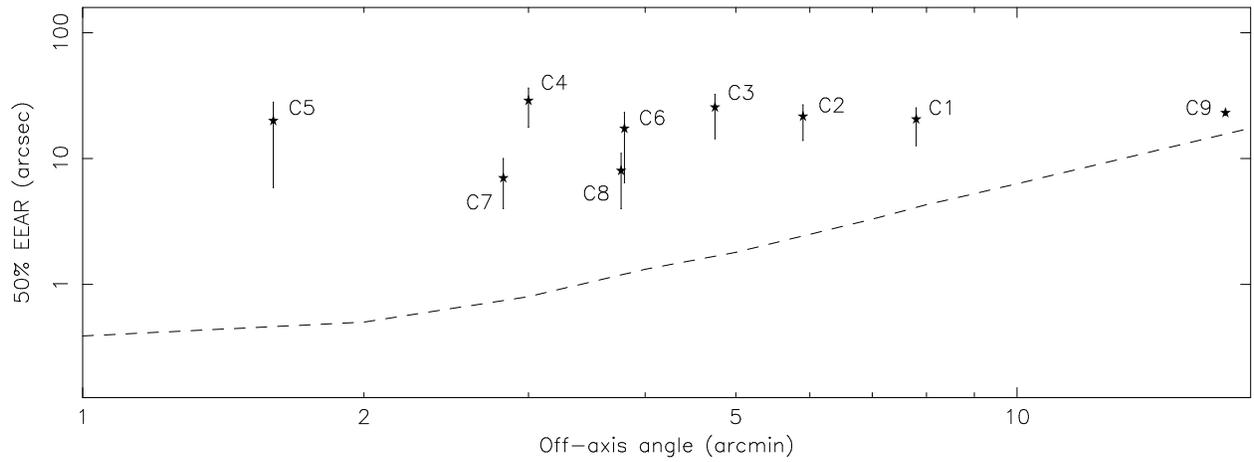}}
  \caption{Angular extension of C1 - C9. The dashed line represents the
50\% EEAR of the PSFs at off-axis angles. Background uncertainties (3 $\sigma$)
are included in the error bars.
    \label{fig:img:smo}}
\end{figure}

\clearpage

\begin{figure*}[htb]
\vspace{-2cm}
  \centerline{\includegraphics[height=1.3\linewidth]{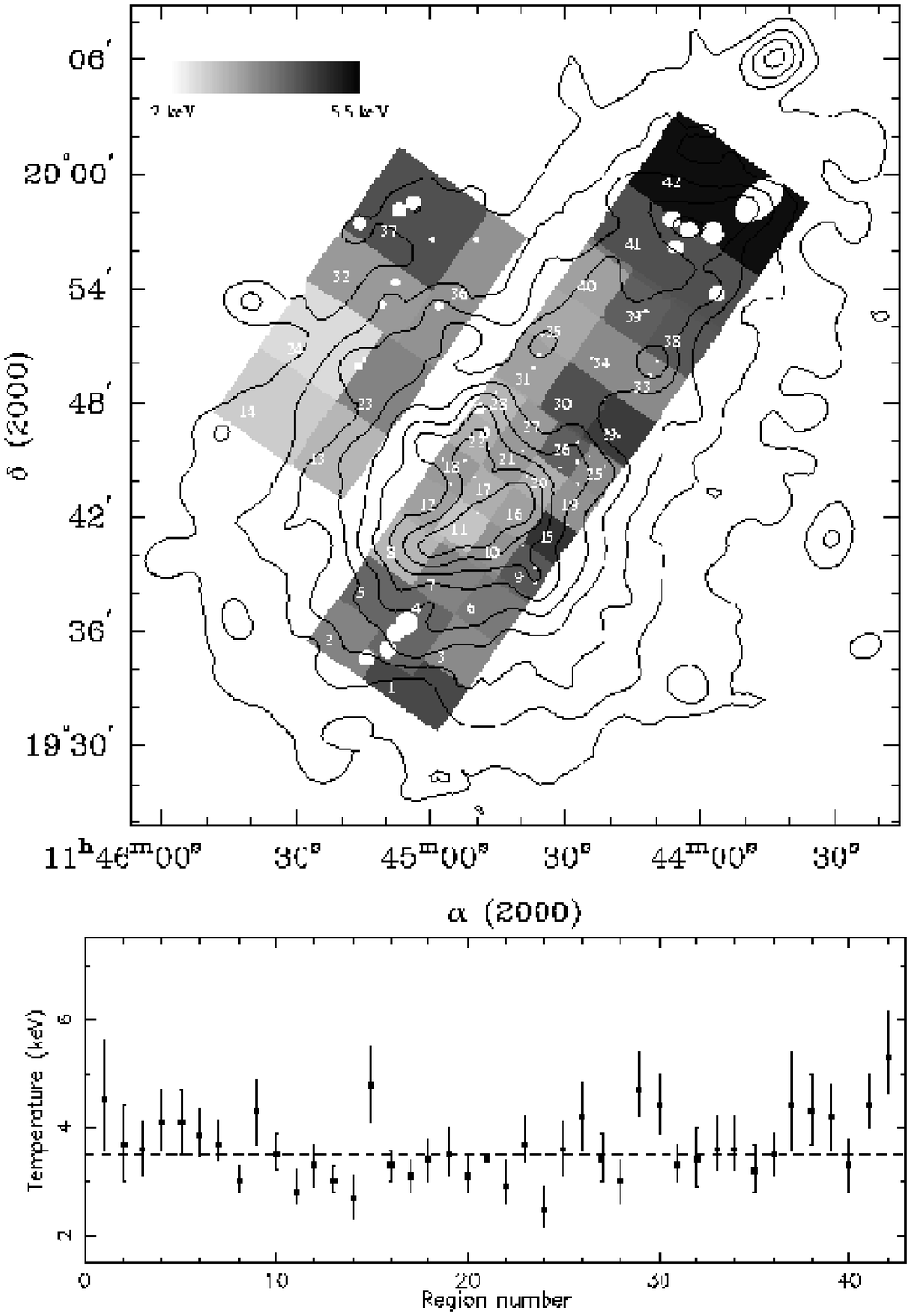}}
\vspace{-0.15cm}
  \caption{The \chandra\ temperature map of A1367. {\bf Upper}: the \chandra\
temperature map overlaid on \rosat\ PSPC contours. The regions with point
sources and the extended features associated with galaxies (C7 - C9)
were excluded. The gaps between chips are generally filled by the average
temperatures of the surrounding regions. {\bf Lower}: the measured temperatures
with errors in 42 regions composing the temperature map. The dashed
line indicates the average temperature in the field - 3.5 keV.
    \label{fig:img:smo}}
\end{figure*}

\clearpage

\begin{figure*}[htb]
  \centerline{\includegraphics[height=0.8\linewidth,angle=270]{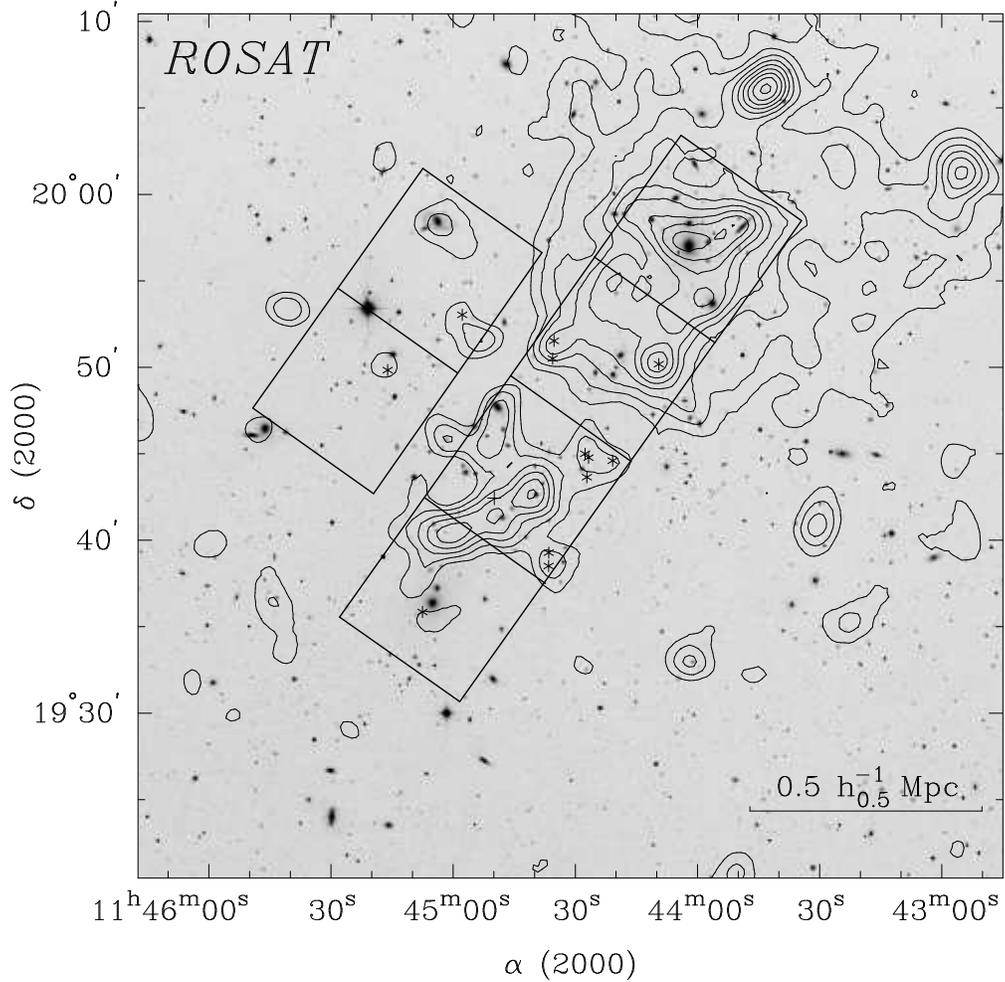}}
  \caption{PSPC residuals after subtracting a spherically symmetrical component
centered at the geometric center of the SE subcluster (the cross).
The component is represented by the best-fit $\beta$-model of the outer parts
of the SE subcluster. The residuals were smoothed by a gaussian with 45$''$
$\sigma$. The contour levels are linearly from 15\% to 95\% of the maximal.
The NW subcluster shows a somewhat disturbed morphology. The left small-scale
features are generally point sources. We mark the positions of some bright
\chandra\ point sources with asterisk. The FOV of the \chandra\ observation
is shown by solid lines.
    \label{fig:img:smo}}
\end{figure*}

\clearpage

\begin{figure}
  \centerline{\includegraphics[height=1.0\linewidth,angle=270]{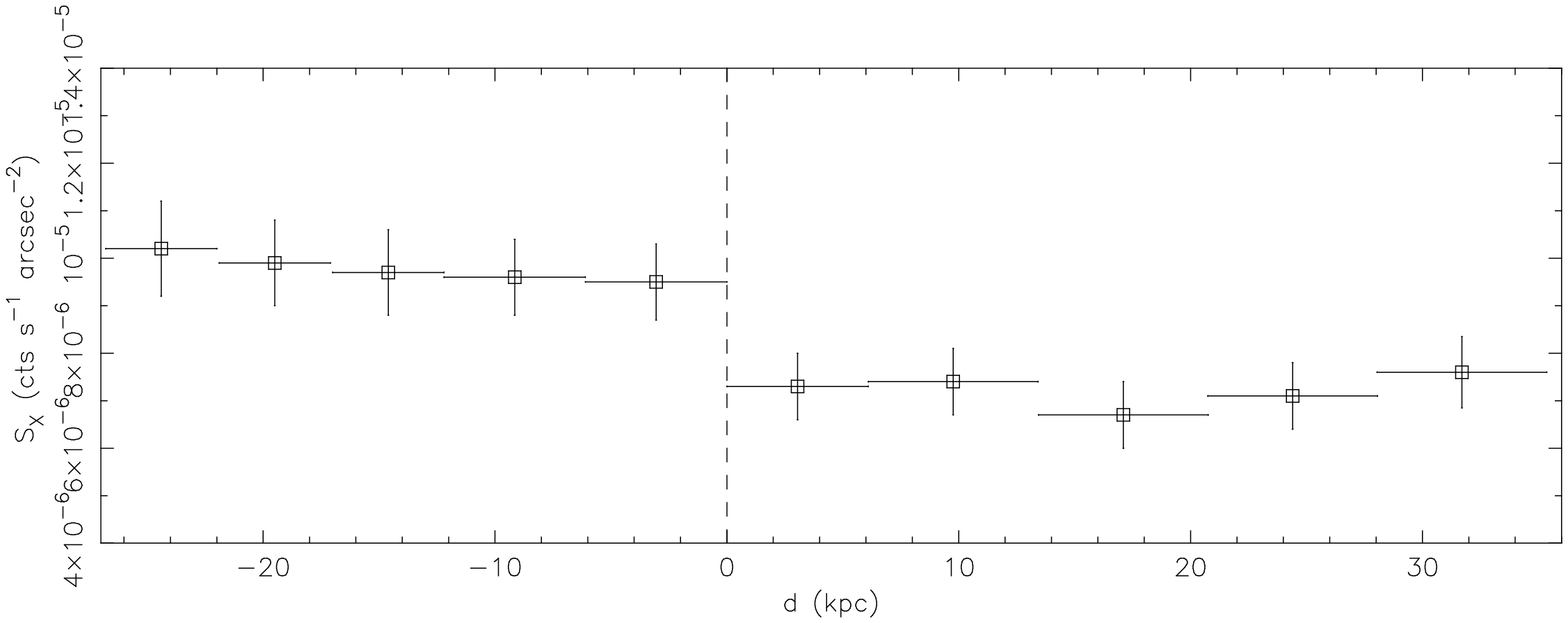}}
  \caption{The 0.5 - 5 keV surface brightness profile across the discontinuity
feature at the southern end of the ridge. The regions chosen to measure the
source brightness are shown in Fig. 3.
    \label{fig:img:smo}}
\end{figure}

\clearpage

\begin{figure}
 \centerline{\includegraphics[height=0.7\linewidth,angle=270]{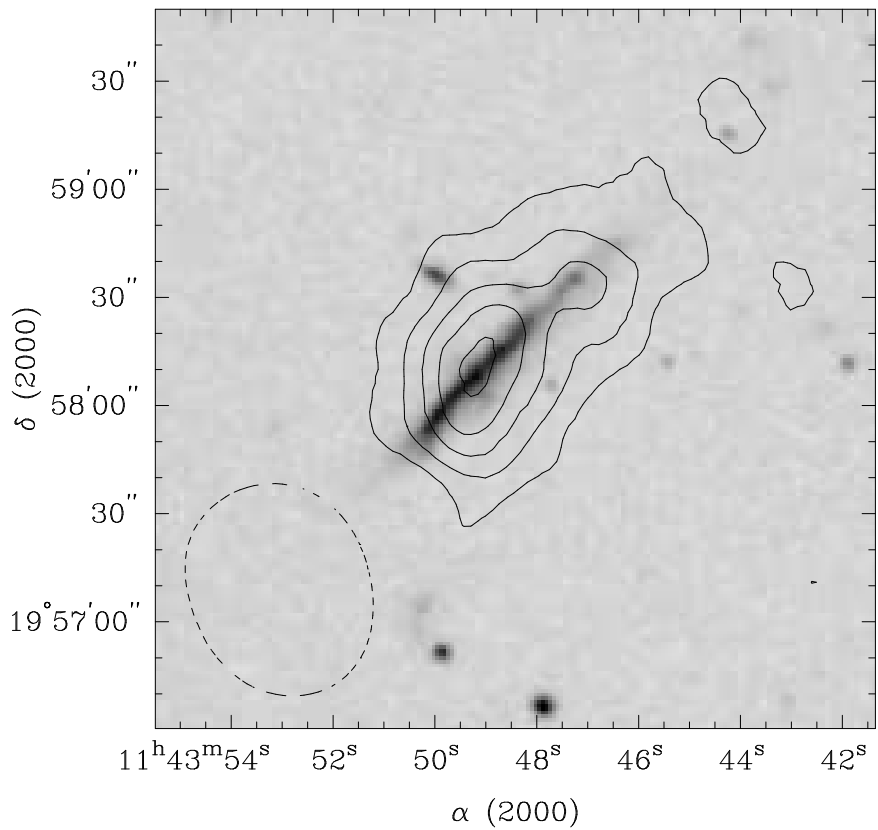}}
 \caption{The X-ray contours of UGC 6697 (smoothed by a gaussian with 7.5$''$
$\sigma$) overlaid on DSS II image.
The surrounding cluster background has been subtracted. The contour levels are
15\%, 35\%, 55\%, 75\%, 95\% of the maximum. The dashed line at the
lower left represents the size of the local PSF (15\% of the maximum in 1.2
keV). The PSF is also smoothed by a gaussian with 7.5$'' \sigma$.
UGC 6697 is clearly resolved in this observation and elongated along optical
orientation.
    \label{fig:img:smo}}
\end{figure}

\clearpage

\vspace{-1cm}
\begin{center}
TABLE 1

{\sc \chandra\ Extended Sources$^{\rm a}$}

{\scriptsize \begin{tabular}{c|c|c|c|c|c|c|c|c|c|c}
\hline \hline
 \# & R.A.$^{\rm b}$ & DEC.$^{\rm b}$ & Size$^{\rm c}$ & S/N$^{\rm d}$ & Counts$^{\rm e}$ & T$^{\rm f}$ & L$_{x}^{\rm g}$ & $\bar{n_{\rm p}}$$^{\rm h}$ & M$_{gas}$$^{\rm i}$ & Comment$^{\rm j}$ \\
 & (J2000) & (J2000) & ($''$) & ($\sigma$) & & (keV) & (10$^{41}$ ergs s$^{-1}$) & (10$^{-3}$ cm$^{-3}$) & (10$^{9}$ M$_{\rm \odot}$) & \\\hline

C1 & 11:45:00.4 & 19:40:32 & 21$^{+5}_{-8}$ & 6.7 & 246$\pm$33 & (2.3) & 1.1 & 1.8 & 3.6 & Gp21 \\

C2 & 11:44:51.5 & 19:40:44 & 22$^{+4}_{-8}$ & 7.2 & 335$\pm$42 & (2.3) & 1.1 & 1.8 & 4.0 & part of Gp17\\

C3 & 11:44:48.3 & 19:41:52 & 25$^{+6}_{-11}$ & 5.0 & 252$\pm$48 & (2.3) & 0.9 & 1.5 & 5.2 & Gp17\\

C4 & 11:44:41.8 & 19:42:48 & 28$^{+7}_{-12}$ & 6.3 & 332$\pm$47 & (2.3) & 1.0 & 1.4 & 6.4 & Gp14\\

C5 & 11:44:40.3 & 19:44:07 & 19$^{+9}_{-13}$ & 4.3 & 165$\pm$34 & (2.3) & 0.5 & 1.6 & 2.4 & \\

C6$^{\rm k}$ & 11:44:52.3 & 19:45:23 & 17$^{+6}_{-10}$ & 4.0 & 127$\pm$30 & (2.3) & 0.4 & 1.5 & 1.6 & \\

C7$^{\rm l}$ & 11:44:47.5 & 19:46:21 & 7$^{+3}_{-4}$ & 5.4 & 124$\pm$25 & 0.3 - 0.8 & 0.3 & 4.6 & 0.3 & Gp18*, NGC 3860 B\\

C8$^{\rm l}$ & 11:44:49.0 & 19:47:41 & 8$^{+3}_{-4}$ & 5.1 & 112$\pm$22 & 0.5 - 0.9 & 0.1 & 4.0 & 0.2 & Gp18, NGC 3860\\

C9$^{\rm l}$ & 11:43:48.7 & 19:58:10 & 23$\pm$1 & 18.9 & 803$\pm$32 & 0.6 - 0.9 & 1.8 & 12.4$^{\rm m}$ & 4.5$^{\rm m}$ & Gp4, UGC 6697\\

\hline\hline
\end{tabular}}
{\footnotesize \begin{flushleft}
\leftskip 10pt
$^{\rm a}$ Detected in the \chandra\ image in the 0.5 - 5 keV energy
band. The sizes, signal-to-noise ratios, and net counts of the features
were also derived based on the image in the 0.5 - 5 keV energy band.\\
$^{\rm b}$ The positions were determined by the peaks of those extended features
in the smoothed image and wavelet reconstructed image. The uncertainty is $\sim$ 5$''$.\\
$^{\rm c}$ Here source sizes are defined as 50\% EEAR, deriving from the
radial profiles of sources. Three $\sigma$ background uncertainties have
been included in the errors. The regions that we used to measure the
significance and net counts, as well as extracting spectra, are
circles with radii of twice 50\% EEAR measured.\\
$^{\rm d}$ We define S/N as S / $\sqrt{(S+B)}$, where S is the signal and B is the background. Background is taken from the surroundings. The detection threshold is set at 4 $\sigma$. The uncertainties of S/N, from the uncertainties
on the background, are less than 10\%.\\
$^{\rm e}$ The net counts are measured from apertures with sizes of twice
50\% EEAR listed on the left. Background is taken from the immediate
surroundings. The errors listed are the uncertainty from 1$\sigma$ errors of
the background.\\
$^{\rm f}$ Temperatures inside the brackets are fixed. 2.3 keV is the best-fit
temperature of integrated C1 - C5 spectrum (the 90\% confidence regime is 1.8
- 3.7 keV). For simplicity, we assumed a same spectrum of C6 to C1 - C5 since
it is hard to constrain its spectrum. For C7 - C9, since
they are relatively cool, their temperatures can be constrained better.\\
$^{\rm g}$ In the 0.5 - 5 keV energy band. Except C6, uncertainties on L$_{x}$
from temperature and abundance are at most 20\%. However, the uncertainties from
the background can reach up to 35\%. For C6, the uncertainty can be
as large as 50\%.\\
$^{\rm h}$ Estimated from the best-fit normalization and assumed a sphere with
constant density. The uncertainties from background and the fits can be up to
30\%. For C6, the density can be changed by a factor of 2 under different
assumption.\\ 
$^{\rm i}$ Except C6, uncertainties can be up to 30\%. For C6, the gas mass 
can be changed by a factor of 2 under different assumption.\\
$^{\rm j}$ Gp\# means the PSPC source \# in G95.\\
$^{\rm k}$ There is a detector feature across C6 so that it is the most poorly
determined one.\\
$^{\rm l}$ To derive the luminosities, densities and gas masses of C7 - C9,
their redshifts rather than the cluster redshift were used. Since the diffuse
X-rays from C7 - C9 may not all come from their thermal halos, the listed
densities and gas masses may be over-estimates.\\
$^{\rm m}$ Since C9 is far from the aimpoint, it is hard to constrain the
physical size of C9. Here we just assumed an optical size of
1.6$'\times$0.5$'$.\\
\end{flushleft}}
\end{center}

\newpage
\begin{center}
TABLE 2

{\sc Extended Sources claimed by previous work$^{\rm a}$}

{\scriptsize \begin{tabular}{c|c|c|c|c|l}
\hline \hline
 \multicolumn{3}{c}{Source ID$^{\rm b}$}  \vline &  \multicolumn{2}{c}{L$_{\rm x}$ (10$^{41}$ ergs s$^{-1}$)$^{\rm f}$} \vline & \mbox{ }\mbox{ }\mbox{ }\mbox{ }\mbox{ }\mbox{ }\mbox{ }\mbox{ }\mbox{ }\mbox{ }\mbox{ }\mbox{ }\mbox{ }\mbox{ }\mbox{ }\mbox{ }\mbox{ }\chandra\ result$^{\rm g}$ \\ \cline{1-5}
B83$^{\rm c}$ & G95$^{\rm d}$ & L98$^{\rm e}$ & B83 & G95 & \\\hline

Be2 & Ge7 (p), Gp10 (p) & Lp26 (p) & - & -, - & bright point source 4 (50\% EEAR of PSF: 4$''$) \\

Be3 & Ge9, - & - & - & 1.3, - & No detection, 3$\sigma$ upper limit: 1.2$\times$10$^{40}$ ergs s$^{-1}$ \\

Be4 & -, - & - & 1.5 & -, - & No detection\\

Be5 & Ge14, Gp11* & - & 3.7 & 0.95,  & bright point source 1 and fainter one (CGCG 097-109) \\

Be6 & Ge17, Gp12 & Lp10 & - & 2.2, 1.8 & bright point source 2 and 3\\

Be7 & Ge19, - & - & 4.6 & 1.9, - & No detection, 3$\sigma$ upper limit: 1.8$\times$10$^{40}$ ergs s$^{-1}$ \\

Be8 & Ge22, Gp18 (p) & Lp16 & 2.9 & 2.0, - & bright nucleus source and extended feature of NGC 3860 (C8) \\

Be9 & Ge21, Gp17 & - & 6.5 & 1.3, 7.7 & extended feature C3, part of Gp17 may include C2\\

Be12 & -, - & - & - & -, - & No detection\\

Be13 & Ge16, - & - & 3.7 & 1.5, - & No detection, 3$\sigma$ upper limit: 1.5$\times$10$^{40}$ ergs s$^{-1}$ \\

Be14 & Ge26 (p) & - & - & -, - & No detection\\

Be15 & -, - & - & 3.5 & -, - & No detection\\

- & -, Gp14 & Lp11 & - & -, 3.6 & extended feature C4 \\

- & -, Gp18*$^{\rm h}$ & - & - & -, 6.9 & extended feature C7 (NGC 3860B)\\

- & -, Gp21 & - & - & -, 1.7 & extended feature C1 \\

- & Ge25 (p), Gp23 & Lp9 (p) & - & -, 2.4 & NGC 3861 and 3861B (all point-like; 50\% EEAR of PSF: 12$''$)\\

- & -, Gp27 & - & - & -, 0.74 & No detection, 3$\sigma$ upper limit: 1.6$\times$10$^{40}$ ergs s$^{-1}$ \\

- & -, Gp4 (p) & Lp7 & - & -, - & UGC 6697 (C9), extended along the optical elongation\\

\hline\hline
\end{tabular}}
{\footnotesize \begin{flushleft}
\leftskip 20pt
$^{\rm a}$ Only those previously claimed extended sources in the field are included (18 of 26) and unless pointed out, all sources are extended. B83 means Bechtold et al. (1983); G95 means Grebenev et al. (1995); L98 means Lazzati et al. (1998).\\
$^{\rm b}$ `-' means no detection, and `(p)' means point source.\\
$^{\rm c}$ Source number in B83 (based on EHRI data); the first letter of the source ID indicates the author and `e' means EHRI. The number represents the order of the source in Table 4 of B83.\\
$^{\rm d}$ Source number in G95 (based on PSPC data and same EHRI data as those used in B83); similarly, the first letter of the source ID represents the author, while `e' means EHRI and `p' means PSPC. The number refers to that in Table 1 and 3 of G95.\\
$^{\rm e}$ Source number in L98 (based on the same PSPC data as those used in G95); the number come from Table 2 of L98.\\
$^{\rm f}$ In the 0.5 - 5 keV energy band. We converted the luminosities listed in B83 and G95 from their energy bands to the 0.5 - 5 keV band and use h=0.5. `-' means no value was provided in the paper.\\
$^{\rm g}$ The source number in Fig. 2a; for the extended features confirmed by \chandra\ (sources represented by 'c'), we list the results in Table 1. The 3$\sigma$ upper limit (0.5 - 5 keV) depends on the assumed spectrum and source size. Here we assumed 1 keV thermal plasma (MEKAL) with a solar abundance (as G95) and used the sizes determined by G95. For such kind of sources with typical sizes 0.5$'$ - 1.5$'$, the 3$\sigma$ upper limits are 1 - 4 $\times$10$^{40}$ ergs s$^{-1}$ (0.5 - 5 keV), which are also dependent on positions on the detector. For B83 sources not recognized as extended sources in both G95 and L98, we do not list the 3$\sigma$ upper limits.\\
$^{\rm h}$ Gp18* was attributed to Be7/8 in G95, but they are different based on the published positions.\\
\end{flushleft}}
\end{center}

\clearpage

\begin{center}
TABLE 3

{\sc Member galaxies detected as X-ray extended features$^{\rm a}$}

{\scriptsize \begin{tabular}{c|c|c|c|c|c|c|c|c|c}
\hline \hline
 Name  & Type & Velocity & Log (L$_{\rm B}$/L$_{\rm \odot}$) & Radio$^{\rm b}$ & Infrared & X-ray$^{\rm c}$ & \#$^{\rm d}$ & Total mass$^{\rm e}$ & Binding mass$^{\rm f}$ \\
   &  & (km/s) &  &  &  &  &  & (10$^{11}$ M$_{\rm \odot}$) & (10$^{11}$ M$_{\rm \odot}$) \\\hline

NGC 3860 & Sa & 5595$\pm$8 & 10.72 & 76 & IRAS 11422+2003 & Ge22, Gp18 & 1, C8 & 3.7 & 1.8 \\

NGC 3860B & Irr & 8293$\pm$9 & 10.53 & 75 & - & Gp18* & 2, C7 & 2.4 & 2.1 \\

UGC 6697 & Irr & 6725$\pm$2 & 11.00 & 58 & IRAS 11412+2014 & Gp4 & 9, C9 & 7 & 3.8 \\

\hline\hline
\end{tabular}}
{\footnotesize \begin{flushleft}
\leftskip 10pt
$^{\rm a}$ The velocity and L$_{\rm B}$ of the galaxies are from NED.\\
$^{\rm b}$ The source number in Gavazzi \& Contursi (1994)\\
$^{\rm c}$ The source number in G95 \\
$^{\rm d}$ The source number in Fig. 2a and Table 1\\
$^{\rm e}$ The total mass is estimated by assuming M/L = 7(M/L)$_{\rm \odot}$.\\
$^{\rm f}$ The binding mass needed to trap the thermal halos observed.
It is estimated by the expression in $\S$7. The sizes of sources are
obtained from Table 1 and their temperatures are assumed to be 0.75 keV.
All estimated binding masses are less than the estimated total masses.\\
\end{flushleft}}
\end{center}

\end{document}